\documentclass[%
 reprint,
superscriptaddress,
 amsmath,amssymb,
 aps,
 prl,
]{revtex4-1}
\usepackage{bbm}
\usepackage{graphicx}
\usepackage{dcolumn}
\usepackage{bm}
\usepackage{braket}
\usepackage{amsmath}
\usepackage{multirow}
\usepackage{graphicx}
\usepackage{comment}
\usepackage[colorinlistoftodos]{todonotes}
\usepackage[colorlinks=true, allcolors=blue]{hyperref}

\begin{document}

\title{Comparison of QAOA with Quantum and Simulated Annealing}
\author{Michael Streif}
\affiliation{Data:Lab, Volkswagen Group, Ungererstra{\ss}e 69, 80805 M{\"u}nchen, Federal Republic of Germany,}
\affiliation{Physikalisches Institut, Universit{\"a}t Freiburg, Herrmann-Herder-Stra{\ss}e 3,
79104 Freiburg, Federal Republic of Germany}
\author{Martin Leib}
\affiliation{Data:Lab, Volkswagen Group, Ungererstra{\ss}e 69, 80805 M{\"u}nchen, Federal Republic of Germany,}
\date{\today}

\begin{abstract}
We present a comparison between the Quantum Approximate Optimization Algorithm (QAOA) and two widely studied competing methods, Quantum Annealing (QA) and Simulated Annealing (SA). To achieve this, we define a class of optimization problems with respect to their spectral properties which are exactly solvable with QAOA. In this class, we identify instances for which QA and SA have an exponentially small probability to find the solution. Consequently, our results define a first demarcation line between QAOA, Simulated Annealing and Quantum Annealing, and highlight the fundamental differences between an interference-based search heuristic such as QAOA and heuristics that are based on thermal and quantum fluctuations like SA and QA respectively.
\end{abstract}

\pacs{Valid PACS appear here}% PACS, the Physics and Astronomy
\maketitle

\paragraph{Introduction --}
The seminal developments of Shor's and Grover's algorithm, that showed a provable exponential and polynomial speedup with respect to their classical counterparts respectively, sparked the decades-long run to build a quantum computer. First quantum computing devices with tens of noisy qubits, so called Noisy Intermediate-Scale Quantum (NISQ) devices have already been built \cite{barends2016digitized,dicarlo2009demonstration,debnath2016demonstration,reagor2018demonstration}. However, to outperform todays most powerful classical computers, Shor's and Grover's algorithms require a fully error-corrected device with of the order of $10^5$ qubits \cite{jones2012layered}. Finding an algorithm for NISQ devices which is superior with respect to its fastest known classical counterpart is therefore the next important step for useful quantum computing. To achieve this, a deep understanding of relations between various quantum algorithms is essential.

Hybrid quantum classical algorithms, which are parameterized quantum circuits that are optimized in a classical learning loop, are generally believed to be the strongest candidates in the NISQ era. Among these are algorithms such as the Variational Quantum Eigensolver (VQE) \cite{peruzzo2014variational} for quantum chemistry calculations, the Quantum Neural Networks (QNN) \cite{farhi2018classification,grant2018hierarchical} for machine learning tasks and the Quantum Approximate Optimization Algorithm (QAOA) \cite{farhi2014quantum}. QAOA can be used to solve combinatorial optimization problems such as MaxCut \cite{wang2018quantum}, Max E3LIN2 \cite{2014arXiv1412.6062F} and generative machine learning tasks such as sampling from Gibbs states \cite{verdon2017quantum}. Interestingly there also exist QAOA versions of Shor's number factoring algorithm \cite{anschuetz2018variational}, and Grover's problem of searching an unstructured database \cite{jiang2017near} that substantially reduce the number of gates with respect to their counterparts for fully error-corrected quantum computers. Moreover it has been shown that there is no efficient classical algorithm that can simulate sampling from the output of a QAOA circuit \cite{farhi2016quantum}.

All the above mentioned algorithms run on gate based quantum computers. However, quantum annealers present an alternative way to enable quantum-enhanced information processing, with the same areas of application, combinatorial optimization and generative learning. There already exists an extensive body of research separating the strengths and weaknesses of Quantum Annealing (QA) and its classical counterpart Simulated Annealing (SA). Classes of problems have been identified that are either tailored \cite{mandra2018deceptive,denchev2016computational} or randomly generated and post-selected \cite{katzgraber2015seeking} to show a quantum speedup of QA, on existing hardware.  

In the present work we add QAOA to this framework of comparisons. We identify a set of problems based on their spectral features which can be solved exactly with at most in the problem size polynomially growing number of gates in the QAOA circuit. Among these there are problems that can not cannot be solved with neither QA nor SA, which we corroborate with their overlap distribution. We further show that for these problem instances there exists an efficient classical algorithm that can find the solution. Therefore, our results provide us with a rich understanding of the nature of the algorithm and show how interference effects separate QAOA from Simulated and Quantum Annealing.

The present article is organized as follows: First we shortly recapitulate the QAOA algorithm. Then we identify the instances that can be solved exactly with one-block QAOA and afterwards show how to construct the corresponding classical problem Hamiltonian. Subsequently we find the subclass of instances that are hard to solve with SA or QA. And finally we show how the fully trained QAOA circuit does not build up entanglement between all possible bipartitions of the qubit register for the above mentioned instances and describe a classical algorithm that can find the solution of these instances. We conclude with discussion and outlook.

\paragraph{The Quantum Approximate Optimization Algorithm (QAOA) --}\label{sec:qaoa}
The Quantum Approximate Optimization Algorithm (QAOA) by Farhi et al.~\cite{farhi2014quantum} is a variational wavefunction ansatz with the goal to find an upper bound of the ground state energy, $E_g$, of a Hamiltonian $H_\mathrm{P}$ which is diagonal in the computational basis, the product states of the eigenstates of $\sigma_z^{(i)}$,
\begin{align}
   E_g= \min_{\Vec{\beta},\Vec{\gamma}}\braket{\Psi(\Vec{\beta},\Vec{\gamma})|H_\mathrm{P}|\Psi(\Vec{\beta},\Vec{\gamma})}.
    \label{minimize}
\end{align}
Various combinatorial optimization problems can be encoded in the ground state of such diagonal Hamiltonians \cite{lucas2014ising}. This way QAOA can be used to solve combinatorial optimization problems. 
The ansatz for the variational wavefunction $\ket{\Psi(\Vec{\beta},\Vec{\gamma})}$ is inspired by the quantum annealing protocol where a system is initialized in an easy to prepare ground state of a local Hamiltonian $H_\mathrm{X}=\sum _i\sigma_x^{(i)}$ which is then slowly transformed to the problem Hamiltonian $H_\mathrm{P}$ \cite{kadowaki1998quantum}. The QAOA variational wavefunction resembles a trotterized version of this procedure,
\begin{align}
\ket{\Psi(\Vec{\beta},\Vec{\gamma})}=\mathrm{e}^{-\mathrm{i}\beta_pH_\mathrm{X}}\mathrm{e}^{-\mathrm{i}\gamma_pH_\mathrm{P}}\dots\mathrm{e}^{-\mathrm{i}\beta_1H_\mathrm{X}}\mathrm{e}^{-\mathrm{i}\gamma_1H_\mathrm{P}}\ket{+}\,,
\label{finalstate}
\end{align}
where the starting state $\ket{+}$ is the product state of eigenstates of $\sigma_x$ with eigenvalue $1$, $\ket{+} = \prod_i (\ket{0}_i + \ket{1}_i)/\sqrt{2}$ which is simultaneously the superposition of all computational basis states.
In contrast to a trotterized version of QA, the parameters $\Vec{\beta}$ and $\Vec{\gamma}$ are adjusted in a classical learning loop to minimize the objective function Eq.~(\ref{minimize}). Various types of outer learning loops have been used thus far ranging from brute force grid search \cite{farhi2014quantum} to gradient based methods \cite{guerreschi2017practical} and recently methods inspired by supervised machine learning where the parameters $\Vec{\beta}$ and $\Vec{\gamma}$ were trained on random samples of combinatorial optimization problems and afterwards kept fixed to solve instances not seen during training of the same combinatorial optimization problem \cite{crooks2018performance,2018arXiv181204170B}.

\paragraph{Spectral conditions for deterministic QAOA --}\label{sec:spectralprops}
In the following we derive conditions for the spectrum of the problem Hamiltonian $H_\mathrm{P}$ such that a one-block version of QAOA ($p=1$) succeeds exactly, i.e. we consider a deterministic version of QAOA where we not only strive to minimize the ground state energy, cf. Eq.~(\ref{minimize}), but search for optimal values of $\beta$ and $\gamma$ such that we find perfect overlap, $|\braket{t|\Psi(\beta, \gamma)}|=1$. Here $\ket{t}$
is the target state of a generic $N$-qubit Hamiltonian that is diagonal in the computational basis, $H_\mathrm{P}=\mathrm{diag}\left(E_1,E_2,\dots,E_{2^N}\right)$. The target state could in general encode the solution to a combinatorial optimization problem of interest. The overlap of the variational wavefunction with the target state can be reformulated by using the structure of $H_\mathrm{X}$ and $H_\mathrm{P}$, 
\begin{multline}
\braket{t|\mathrm{e}^{-\mathrm{i}\beta H_\mathrm{X}}\mathrm{e}^{-\mathrm{i}\gamma H_\mathrm{P}}|+}\\
=\frac{1}{\sqrt{2^N}} \sum\limits_{l=1}^{2^N} \mathrm{e}^{-i(\gamma E_l + \frac{\pi}{2}\Delta_t(l))}\cos(\beta)^{N-\Delta_t(l)}\sin(\beta)^{\Delta_t(l)}\,.
\label{overlap} 
\end{multline}
Here, $\Delta_t(l)$ is the Hamming distance of the computational state $\ket{l}$ w.r.t. the target state $\ket{t}$, i.e. the number of spin flips required to change the state $\ket{l}$ to the state $\ket{t}$. This overlap is a sum of $2^N$ complex numbers, where the magnitude of each summand is maximally $1/2^N$ for $\beta=\pi/4$. A deterministic version of the QAOA algorithm demands that this sum adds up to a complex number with a magnitude equal to $1$. Therefore, if there is a perfect solution it is only achievable with $\beta=\pi/4$.
To get the deterministic solution however we have to additionally demand that the phase $c$ of all complex numbers is the same,
\begin{equation} 
    (\gamma E_l + \frac{\pi}{2} \Delta_t(l)) \, \text{mod}\, 2 \pi = c \qquad \forall\hspace{1mm}l \in \left\{1,2,\dots, 2^N \right\}.
    \label{eq:spectralconditions}
\end{equation}
These conditions define a subclass of problem Hamiltonians which can be solved exactly with one-block QAOA.

\paragraph{Corresponding spin glass systems --}
To convert the diagonal Hamiltonian to a quantum circuit, we reformulate it as spin glass Hamiltonian,

\begin{align}
H=&\sum_{i_1}^N h_{i_1} \sigma_z^{(i_1)} + \sum_{i_1,i_2}^N J_{i_1i_2}\sigma_z^{(i_1)}\sigma_z^{(i_2)}\nonumber\\&+ \sum_{i_1,i_2,i_3}^N J_{i_1i_2i_3}\sigma_z^{(i_1)} \sigma_z^{(i_2)} \sigma_z^{(i_3)} + \dots
\label{gspinglass}
\end{align}
given in terms of their on-site fields ($h_i$) and up to $k$-local interactions ($J_{i_1i_2}, J_{i_1i_2i_3}, \dots,J_{i_1i_2i_3\dots i_k}$), that fulfill the requirements of the instances found above. To implement the evolution generated by this Hamiltonian, we transform every term to a $k$-qubit gate. To fulfill the above defined conditions on the spectrum, it is necessary to group the states according to their Hamming distance with respect to the target state $\ket{t}$ we would like to find with QAOA. We construct the spin glass Hamiltonian with the help of the term
\begin{equation}
     \sum\limits_i^N t_i \sigma_z^{(i)}= N - 2 \tilde{\Delta}_t \,,
\end{equation}
where $\sigma_z^{(i)}$ is the Pauli matrix acting on qubit $i$ and $\tilde{\Delta}_t$ is the Hamming distance operator defined by the eigenstates given by the computational basis states and the eigenvalues given by the Hamming distance of the respective computational basis state and target state $\ket{t}$. We decompose the spin glass Hamiltonians for our instances into two parts,
\begin{equation}\label{eq:sg-rep}
    H_{\mathrm{SG}} = \frac{\pi}{4}\sum\limits_i^N t_i \sigma_z^{(i)} + H_{2 \pi}\,. 
\end{equation}
The first term fixes the conditions given in Eq.~(\ref{eq:spectralconditions}) and the second term $H_{2 \pi}$ is an arbitrary spin glass with the sole condition that all eigenvalues are multiples of $2\pi$, which can be adjusted for any spin glass by rescaling of the energies. This means that we can add a watermark state $\ket{t}$ to every arbitrary spin glass such that QAOA deterministically creates this state which can be any state of the respective spin glass, not necessarily the ground state.

\paragraph{SA/QA-hard instances --}
\label{sec-hardinstances}

Among the above defined instances there are problems that are hard to solve for both QA as well as SA. Both of these methods are heuristics designed to find a state that minimizes the energy of a given spin glass. 

For SA one starts in a random computational basis state and performs a random walk in the configuration space with Metropolis--Hasting updates with the goal to relax to low lying minima of the potential landscape. On the way to the solution, the found energy barriers can be overcome if their height is of the order of the thermal fluctuations or smaller. When cooling down the temperature slowly, in the best case scenario, SA finds the global minimum of the energy landscape.

For QA in comparison a system is initialized in the superposition of all computational basis states and the magnitude of the quantum fluctuations are decreased until the system settles in a minimum of the potential landscape. Tunneling has been proven to be beneficial in this process \cite{denchev2016computational}. Tunneling through a barrier is exponentially suppressed as a function of the barrier width while it is proportional to the inverse of the barrier height. 

QA therefore shows advantages compared with SA for potential landscapes where minima are separated by thin and tall barriers while both heuristics fail for minima separated by tall and wide barriers \cite{katzgraber2015seeking}. We therefore identify the two requirements for instances that are hard to solve for QA and SA: First, the potential landscape should feature a large number of minima separated by wide barriers, where the relevant metric in this case is Hamming distance. Second, only one minimum should be the global minimum  with all other minima separated by an energy scale which is considered to be large enough such that the specific non-optimal minimum cannot be considered to be an acceptable solution to the encoded problem. 

\begin{center}
\begin{figure*}[t!]
\centering
\includegraphics[width=\textwidth]{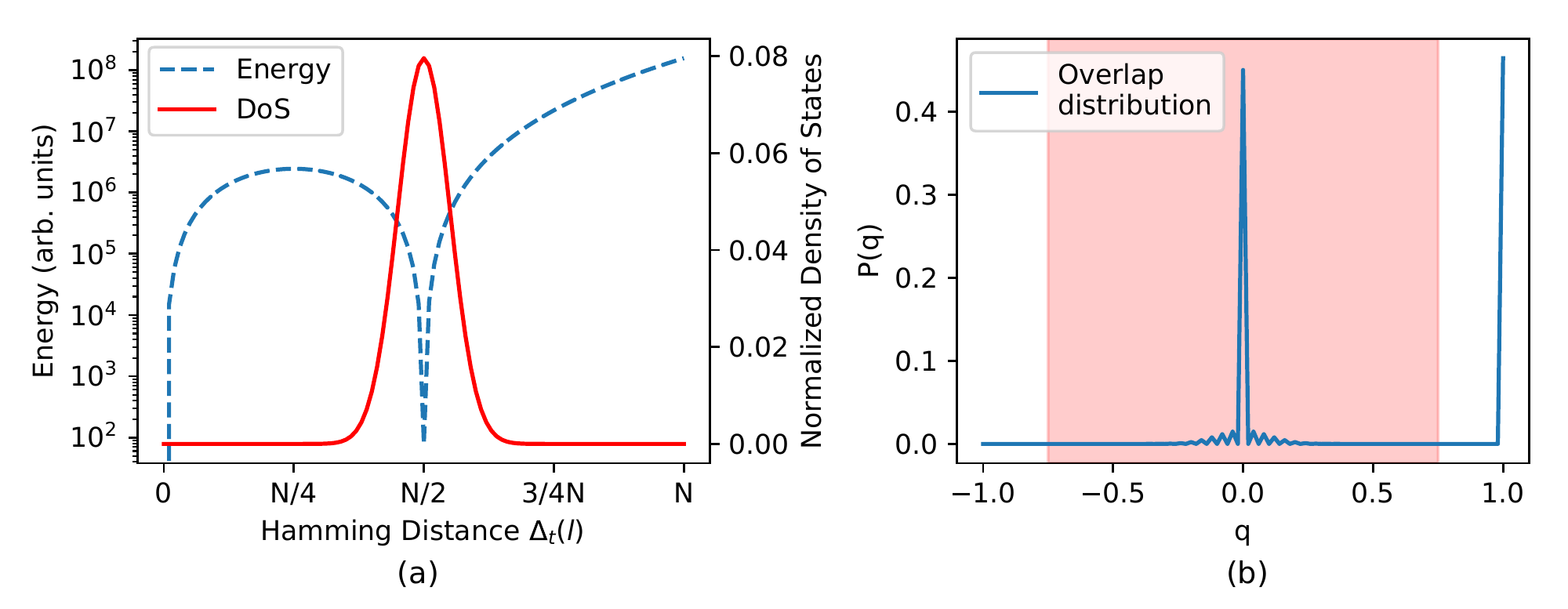}
\begin{minipage}[t!]{1\linewidth}
	\caption{(a) The dashed line shows an artificially constructed energy distribution, which fulfills the spectral conditions given in Eq.~(\ref{eq:spectralconditions}) and employs maximal 4-local interactions. The solid line shows the normalized density of states w.r.t. the Hamming distance to highlight that the energy landscape is dominated by many sub optimal minima. (b) Overlap distribution for the given spectrum. The peak at $q=1$ denotes the overlap of every minimum with itself. The peak at in the red/dark area, however, is the overlap of all suboptimal minima with the global minimum. The position at q=0 means that they are mainly located at a Hamming distance of N/2. Following \cite{katzgraber2015seeking}, the peaks around $q<|0.75|$ indicate that both SA and QA will struggle to find the global minimum. Both plots show numerical data for $N=100$.}
	\label{fig:4local}
\end{minipage}
\end{figure*}
\end{center}

In general, we can generate spin glasses with arbitrary eigenenergies. However, this could lead to $k$-local interactions up to the maximal $N$-locality. This in turn leads to a decomposition of the problem Hamiltonian block in the QAOA algorithm with an exponentially growing number of elementary gates. We therefore add an additional requirement of finite $k$-locality of the spin glass, where $k$ is independent of the size of the problem. The instances we found that fulfill the above requirements with maximal 4-local terms are the following, 
\begin{equation}
    H_{2 \pi}= 2\pi \tilde{\Delta}_t^2 (\tilde{\Delta}_t - (N/2))^2 + H_{2\pi}'\,,
\end{equation}
where $H_{2\pi}'$ is another arbitrary spin glass Hamiltonian with the sole requirement that its eigenenergies are multiples of $2\pi$ and that the interactions may not be greater than 4-local. 
The quartic polynomial in the Hamming distance operator ensures that the target state is also a ground state of the spin glass while at the same time it generates an exponential number $\binom{N}{N/2}$ of minima with Hamming distance $N/2$. These minima are suboptimal because of the first part of Eq.~(\ref{eq:sg-rep}). In Fig.~\ref{fig:4local}~(a), we show the energy distribution as function of the Hamming distance and the density of states w.r.t. the Hamming distance. The density of states visualizes that an exponentially large fraction of random starting points in classical methods will be close to sub optimal minima.

To provide numerical evidence that these constructed instances are hard for both SA and QA and to make contact with the notions introduced in \cite{katzgraber2015seeking}, we calculate their overlap distributions. The overlap distribution is defined as the probability distribution of
\begin{align}
    q=\frac{1}{N}\sum_i^N s_i^{(\alpha)} s_i^{(\beta)}.
\end{align}
defined over two replicas, $\alpha$ and $\beta$, of the system in a thermodynamic state. It is shown that the overlap distribution allows to draw conclusions about the hardness of combinatorial problems for both Simulated Annealing and Quantum Annealing.  Instances with peaks in the overlap distribution for small values of $q$ have been identified as hard to solve for both QA and SA \cite{katzgraber2015seeking}. We calculated the overlap distributions exactly, cf. the Supplementary Material, for the 4-local instances found above where $H_{2\pi}' = 0$. We find perfect alignment of our instances with heuristics found in \cite{katzgraber2015seeking} for hard instances for QA and SA, cf. Fig.~\ref{fig:4local}~(b). 

\paragraph{Classical algorithm --}
The fully trained version of the QAOA circuit for instances with deterministic outcome as defined above, cf. Eq.~(\ref{eq:spectralconditions}), does not build up any entanglement, as can be seen from the following observation,
\begin{equation}
    e^{-\mathrm{i}\left(\frac{\pi}{4}\sum\limits_i t_i\sigma_z^{(i)} + H_{2\pi}\right)} = \prod_i e^{-\mathrm{i} \frac{\pi}{4} t_i \sigma_z^{(i)}}\,,
\end{equation}
which is why every gate in the fully optimized QAOA circuit is local. This suggests that there is an efficient classical algorithm to find solutions for these instances. In the following we present an efficient classical algorithm that can find the target state given oracle access to the energies of computational basis states of the Hamiltonian given in Eq.~\ref{eq:sg-rep}: First, one queries the energy of a random computational basis state. Second, a random spin of the initial state is flipped. The Hamming distance of the resulting state w.r.t. the target state then is either increased or decreased by one. If the Hamming distance is increased  by one, then we know the initial state of the spin was the correct one. If the Hamming distance is decreased by one, then we can leave the spin as is. To see if the Hamming distance was increased or decreased we query the energy for the state with flipped spin and examine this energy modulo $2\pi$. The remainder is the Hamming distance from the target state modulo 4 multiplied by $\pi/4$. This suffices to detect with certainty if the Hamming distance was increased or decreased by one. We repeat the above described method for every spin and are able to find the target state with $N+1$ queries of the oracle. 

\paragraph{Conclusion --}
QAOA can be seen as trotterized version of Quantum Annealing \cite{farhi2014quantum}. However, our results show that QAOA is able to deterministically find the solution of specially constructed optimization problems in cases where both Quantum Annealing and Simulated Annealing fail. Consequently, our results define a first demarcation line between QAOA on one side and SA and QA on the other side. These results highlight the fundamental differences between heuristics designed to find the minimum of potential landscapes such as QA and SA and an interference-based algorithm such as QAOA where all states that are not the target state interfere destructively while only the the amplitudes of the target state add up constructively. 

We moreover showed that there exists an efficient classical algorithm for these instances as suggested by the lack of building up of any entanglement in the trained QAOA circuit. Interestingly, these instances are easy from a computational point of view, however QA and SA are both tricked to find local minima instead of the true solution.

As the found instances define a complete set of the exactly soluble problems for $p=1$, our findings point to intensifying research on probabilistic versions of QAOA or on deterministic versions with $p>1$ for combinatorial optimization problems as well as generative machine learning tasks to find practical applications of QAOA that harness the proven worst-case complexity of QAOA circuits \cite{farhi2016quantum}. Moreover, finding embedding schemes which take advantage of the here presented spectral conditions could help to find more powerful versions of QAOA. Similarly, intensifying research on the impact of interference effects on the computational power of quantum computing is the key to find suitable problems and new algorithms. 
\label{sec-summary}
\paragraph{Acknowledgments --}
The authors would like to thank Eddie Farhi, Jeffrey Goldstone, Masoud Mohseni, Andrea Skolik, Patrick van der Smagt, Andreas Buchleitner and Filip Wudarski for useful comments and enlightening discussions. We thank VW	Group	CIO	Martin	Hofmann, who	 enables	 our	 research.
\bibliographystyle{unsrt}
\bibliography{sample}
\newpage
\onecolumngrid
\appendix
\section{Overlap distribution}\label{supp-mat:overlap}
Here we calculate an exact expression for the overlap distribution in the case where the problem Hamiltonian is defined via the Hamming distance to the target state $\ket{t}$. The overlap is defined by 
\begin{align}
    q=\frac{1}{N}\sum_i^N s_i^{(\alpha)} s_i^{(\beta)},
\end{align}
where $\alpha$ and $\beta$ define two replicas of the system.
The probability distribution for each combination of states of the two replicas is given by the product of two Gibbs distributions at the same temperature. We calculate the probability distribution of $q$ by summing over combinations of states of the two replicas that amount to the same value of $q$,
\begin{align}
    P(q)=\sum_{i,j}^N \mathrm{e}^{-\beta H(s_i)}\mathrm{e}^{-\beta H(s_j)}\delta_{s(i)s(j),qN}.
    \label{overlapsum}
\end{align}
We note that the value $q$ is directly related to the Hamming distance, $\Delta H$, between two computational states, $q=(N-2\Delta H)/N$. For an arbitrary Hamiltonian, evaluating this sum requires exponential many classical resources as there are, in general, exponential many products of different energy pairs. However we consider a problem Hamiltonian that depends solely on the Hamming distance to the target state. Therefore we only have $N+1$ different energies and consequently only $N(N+1)/2$ different Gibbs weight pairs. In the following, we show how to calculate the overlap distribution with polynomial resources in this case.

To find all possibilities to create a certain value of $q$, we have to sum over all possibilities to create a certain distance $\Delta H$ from all possible computational states $s_i$. Let us have a look at an example, where our target state is $\ket{t}=\ket{0}^{\otimes N}$. If the system is in a computational state $l$, $\Delta_t(l)$ qubits are in the $\ket{1}$ state and consequently $N-\Delta_t(l)$ qubits are in the $\ket{0}$ state. Note that we here used $\Delta_t(l)$ as the Hamming distance from the target state w.r.t. to the state $s_i$. We now have to sum over all possible combinations, leaving us with the
\begin{align}
    P(q)=&\sum_{\Delta_t=0}^{N}\sum_{K=0}^{\Delta{H}}\binom{N}{\Delta_t}\binom{N-\Delta_t}{\Delta{H}-K}\binom{\Delta_t}{K} \mathrm{e}^{-\beta E(\Delta_t)}\mathrm{e}^{-\beta E(\Delta_t+\Delta H-2K)}.
\end{align}
This expression has only polynomial many terms and can be computed efficiently. Usually, the overlap distribution is found by doing many simulated annealing runs with different temperatures. In our case, we can set the temperature arbitrarily. We therefore have to make sure to set the temperature such that we do not hit both thermodynamic limits, where i) the ground state is occupied with certainty or ii) all states are equally likely. Therefore we shift the temperature in the regime where both cases i) and ii) do not occur. In this case, the overlap distribution gives insight into the energy landscape of the problem and the capability of QA and SA to solve the problem.
\label{appendixoverlapdist}
\end{document}